# Magnetization-direction-tunable kagome Weyl line

Zi-Jia Cheng[*†1], Ilya Belopolski[*1], Tyler A. Cochran[*1], Hung-Ju Tien[*2], Xian P. Yang[*1], Wenlong Ma[3], Jia-Xin Yin[1], Junyi Zhang[4], Chris Jozwiak[5], Aaron Bostwick[5], Eli Rotenberg[5], Guangming Cheng[6], Md. Shafayat Hossain[1], Qi Zhang[1], Nana Shumiya[1], Daniel Multer[1], Maksim Litskevich[1], Yuxiao Jiang[1], Nan Yao[6], Biao Lian[4], Guoqing Chang[7], Shuang Jia[3], Tay-Rong Chang[†2], M. Zahid Hasan[†1,8]

1. Laboratory for Topological Quantum Matter and Advanced Spectroscopy (B7), Department of Physics, Princeton University, Princeton, NJ, USA
2. Department of Physics, National Cheng Kung University, Tainan, Taiwan
3. International Center for Quantum Materials, School of Physics, Peking University, Beijing, China
4. Department of Physics, Princeton University, Princeton, New Jersey 08544, USA
5. Advanced Light Source, E. O. Lawrence Berkeley National Laboratory, Berkeley, California 94720, USA
6. Princeton Institute for Science and Technology of Materials, Princeton University, Princeton, NJ, USA
7. Division of Physics and Applied Physics, School of Physical and Mathematical Sciences, Nanyang Technological University, 21 Nanyang Link, 637371, Singapore
8. Lawrence Berkeley National Laboratory, Berkeley, CA 94720, USA

**Kagome magnets provide a fascinating platform for a plethora of topological quantum phenomena. Here, utilizing angle-resolved photoemission spectroscopy, we demonstrate Weyl lines with strong out-of-plane dispersion in an A-A stacked kagome magnet $Tb_xGd_{1-x}Mn_6Sn_6$. On the Gd rich side, the Weyl line remains nearly spin-orbit-gapless due to a remarkable cooperative interplay between Kane-Mele spin-orbit-coupling, low site symmetry and in-plane magnetic order. Under Tb substitution, the kagome Weyl line gaps due to a magnetic reorientation to out-of-plane order. Our results illustrate the magnetic moment direction as an efficient tuning knob for realizing distinct three-dimensional topological phases.**

† zijiac@princeton.edu, u32trc00@phys.ncku.edu.tw, mzhasan@princeton.edu

Topological quantum magnets can host electronic structures with Dirac and Weyl points and lines near the Fermi level, which are effective in concentrating Berry curvature in bulk momentum space [1–6]. This large Berry curvature, in turn, has been observed to drive giant anomalous Hall and Nernst effects [7–13], even up to room temperature [14–16], making these exotic magnets promising candidates for magnetic field sensors [10,17], thermoelectric converters [14,18] and Berry curvature memories [19,20]. Manipulation of the global topology and Berry curvature geometry in such magnetic semimetals promises to enable new exotic response and device functionality, but remains experimentally challenging. While magnetic topological insulators have long been manipulated by doping and gating to obtain exotic topological states [21–24], known room temperature topological magnetic semimetals are three-dimensional bulk with comparatively high carrier density. The manipulation of the intrinsic bulk magnetic order offers a new design principle for next-generation field-free topological engineering, but has been little studied [25,26].

The kagome lattice exhibits frustration and two-dimensional Dirac points, suggesting that stacked kagome materials may naturally allow magnetic tunability of electronic topology [6,27–33]. Recently, $R$Mn$_6$Sn$_6$ ($R$ is a rare Earth) has emerged as a versatile family of kagome magnets offering rich magnetic orders with different rare earth element $R$ [34,35], critical temperatures above room temperature and a topological Chern gap arising from a spin-orbit gapped kagome Dirac point [4]. For example, GdMn$_6$Sn$_6$ is an easy-plane ferrimagnet with a Curie temperature $T_C$~440 K [35,36], while TbMn$_6$Sn$_6$ is an easy-axis ferrimagnet with $T_C$ ~ 420 K [35]. Substituting Gd with Tb in alloy Tb$_x$Gd$_{1-x}$Mn$_6$Sn$_6$ can further smoothly tune the magnetic ground state without introducing impurity on Mn kagome layer [[37], also see Figs.S3-4 in [39]], offering an attractive platform for understanding the interplay between magnetic momentum direction and topological electronic structure. Moreover, deviating from ideal quasi-2D kagome model, the low anisotropy in resistivity and highly three-dimensional character of *ab initio* electronic structure calculations [36] suggests a significant interlayer coupling, motivating investigation of three-dimensional topology in $R$Mn$_6$Sn$_6$ emerging from the A-A stacked Mn lattice. Here we employ angle-resolved photoemission spectroscopy (ARPES) to probe the three-dimensional electronic structure of Tb$_x$Gd$_{1-x}$Mn$_6$Sn$_6$, providing the evidence, for the first time, of three-dimensional magnetic nodal lines near the Fermi level ($E_f$) with magnetization-direction tunable spin-orbit coupling (SOC) gap in $R$Mn$_6$Sn$_6$.

Ultraviolet ARPES measurements were performed at beamline 5.2 in the Stanford Synchrotron Light Source (SSRL), beamline 7.0.2 in the Advanced Light Source (ALS) and Bloch beamline in the MAX IV. The energy (angle resolution) was better than 20meV (0.2 degree), respectively. Soft X-ray (SX) ARPES data was collected at the ADRESS beamline at Switzerland Light Source (SLS) [38], with 150meV energy resolution at 650eV photon energy. Samples were cleaved and measured between 10K and 20K. The details of sample preparation, TEM, magnetization characterization and calculation can be found in [39].

GdMn$_6$Sn$_6$ has a layered hexagonal crystal structure (P6/mm) [37]. The unit cell exhibits a stacking pattern R-M-S1-S2-S1-M-R along the c direction, where R, M, S1, S2 represent GdSn$_2$ honeycomb layers, Mn kagome layers, Sn hexagonal layers and Sn honeycomb layers, respectively [Fig. 1(a,b)]. The two clean Mn kagome layers in one unit cell form a simple A-A stacking with interlayer separation close to c/2, in sharp contrast to other well-studied kagome magnets such as $Fe_3Sn_2$ [8], $Mn_3Sn$ [7] and $Co_3Sn_2S_2$ [10]. At low temperature, $GdMn_6Sn_6$ shows in-plane ferrimagnetic order, but upon substitutional Tb doping ($Tb_{0.2}Gd_{0.8}Mn_6Sn_6$) exhibits a magnetic reorientation transition to an out-of-plane easy axis [Figs. 1(c), S2 [39], [37]].

We first examine our $GdMn_6Sn_6$ samples by high-resolution ARPES on the (001) cleaving plane at incident photon energy 140 eV with linear horizontal (LH) polarization and the results are shown in Figs. 1(e,f). The spectra readily reveal a cone dispersion at the $\bar{K}$ point of the surface Brillouin zone (BZ) with crossing point energy $E_D = -(0.04 \pm 0.01)$ eV and Fermi velocity $v_{KK} = (6.7 \pm 0.8) \times 10^4$ m/s$^{-1}$. This Fermi velocity is significantly smaller than that of graphene, but typical among transition metal kagome magnets [4,28], suggesting a Dirac cone state arising from localized 3d electrons on the Mn kagome lattice. Since the system is ferrimagnetic, the electronic structure is generically singly-degenerate and these Dirac cones are spin-polarized.

Having observed the Dirac cone state originating from the two-dimensional kagome sheets, we next consider the evolution of this electronic structure in three-dimensional momentum space. We first acquire an ARPES photon energy dependence to probe the electronic structure at different $k_z$ [Fig. 2(a)]. Surprisingly, for the state indicated by the green arrows, we observe a clear oscillation of intensity with a periodicity of $4\pi/c$, doubling the expected $2\pi/c$ periodicity of the nominal bulk BZ. This oscillation indicates a doubling of the effective BZ, or halving of the effective crystallographic unit cell [Fig. 2(b), in the following we will adopt the notation of doubled BZ for the high symmetry points]. To further explore this BZ doubling effect, we examine the electronic structure at $k_z = 9\frac{2\pi}{c}$ ($hv = 140$ eV) and $k_z = 10\frac{2\pi}{c}$ ($hv = 172$ eV). In the nominal Bz, these two $k_z$ both correspond to Γ, but now in the doubled BZ they correspond to the Γ-K-M and $A_u$-$H_u$-$L_u$ planes, respectively. Zooming in on the FS at $k_z = 9\frac{2\pi}{c}$ shown in Fig 2(c), we observe that the upper branch of the Dirac state, labelled $D_1$, appears as a crescent-moon-shaped pocket centered at $H_u$. Mimicking graphene [40] and FeSn [28], the observed pattern is consistent with sublattice interference of the lattice-driven nodal state and a manifestation of the π Berry phase of the wavefunction [39,41,42]. In contrast, at $k_z = 10\frac{2\pi}{c}$ the circular Dirac pocket increases in size and its distribution of spectral weight changes dramatically [Fig. 2(f)]. Moreover, the K-K' cut exhibits hole-like linearly dispersing bands [Fig. 2(g)]. The full dispersion is well captured by a fit to a minimal kagome model, which is shown as dashed line in Fig. 2(g), suggesting a kagome Dirac crossing point (label $D_2$) at 130 meV above $E_F$. We observe that the Fermi velocity of $D_2$ is nearly twice that of $D_1$, indicating distinct Mn orbital origins for the $D_1$ and $D_2$ kagome Dirac cones [also see Figs. S7-8]. To understand the interplay between the Mn kagome Dirac cones and doubled BZ, we note that the conventional unit cell contains two Mn kagome planes, with almost equal intra-unit-cell distance (4.51Å) and inter-unit-cell distance (4.49Å), naturally halving the effective crystallographic unit cell [left panel in Fig. 2(b)]. As the states with energy near $E_f$ are mainly from Mn 3d orbitals [Fig. S11 [39]], the observed zone-selective behavior of the states can be understood as an interference effect from the electrons emitting from adjacent kagome planes, which carries phase information of the initial states [39]. We further support this interpretation by band unfolded Green's function *ab initio* calculations, which capture both the observed dispersion and spectral weight distribution observed by photoemission for alternating $n$ [Figs. 2(e,h)]. The clear BZ doubling with alternating appearance of distinct Dirac states indicates the bulk nature of the kagome Dirac cones, and further points to a significant coupling between Mn kagome layers, suggesting an emergent three-dimensional magnetic topological state.

To probe the evolution of the kagome Dirac states in the stacking direction, we explore our GdMn$_6$Sn$_6$ samples by bulk sensitive soft-X-ray ARPES(SX-ARPES). We acquire a photon energy dependence to extract the E - $k_z$ dispersion along K-H$_u$ [Fig. 3(a), Fig. S10-11 [39]], which cuts through the D$_1$ kagome Dirac cone. Remarkably, we observe a considerable out-of-plane $k_z$ bandwidth of 120 meV for the kagome Dirac state, with the band top at H$_u$ and the band bottom near H. We further acquire conventional ARPES spectra on the (100) side surface to more directly probe, without photon energy dependence, the E - $k_z$ dispersion through D$_2$ [Fig. 3(b)]. Again we observe a large $k_z$ bandwidth of 230 meV, suggestive of a strongly out-of-plane dispersive kagome Dirac state. The $k_z$ dispersion for both three-dimensional Dirac cone structures quantitatively agrees with our band-unfolded *ab initio* calculations shown in Fig. 3(c). Our $k_z$ resolved photoemission spectra suggest that under interlayer coupling the D$_1$ and D$_2$ kagome Dirac cones independently form emergent three-dimensional magnetic nodal structures with large $k_z$ dispersion.

To more deeply understand the magnetic nodal structure, we examine our *ab initio* calculation along K-H$_u$ without SOC. We find that the two-fold degeneracy of D$_1$ is preserved along the entire K-H$_u$ path of the bulk BZ, forming a magnetic nodal line structure which can be understood as a natural consequence of the A-A stacking of kagome sheets [Fig. 3(c) inset]. Including SOC in our calculation, we find that the induced gap remains extremely small, < 0.5 meV, along the full trajectory of the nodal line [Fig. 4(f)]. By contrast, other kagome magnets , such as Fe$_3$Sn$_2$ [8] and TbMn$_6$Sn$_6$ [4], exhibit a moderate SOC gap of ~30 meV. The small SOC gap can be understood as arising from cooperative interplay between low site symmetry and in-plane magnetic order. First, the low site symmetry lifts the degeneracy of all 3$d$ orbitals, suppressing on-site SOC. The next leading contribution is the inter-site Kane-Mele coupling, $H_{KM} \propto J_z S_z$, where $J_z$ and $S_z$ are the out-of-plane components of the orbital and spin angular momenta, respectively. However, the Kane-Mele SOC term is also strongly suppressed under the in-plane magnetic order of GdMn$_6$Sn$_6$, resulting in negligible $S_z$. The interplay of A-A kagome stacking, low site symmetry and in-plane magnetic order naturally produces a Weyl line with near-two-fold degeneracy in GdMn$_6$Sn$_6$.

Having provided evidence for a kagome Weyl line in GdMn$_6$Sn$_6$, we next explore magnetically tuning this topological electronic structure. We investigate Tb$_{0.2}$Gd$_{0.8}$Mn$_6$Sn$_6$, with out-of-plane magnetic order, by ARPES near H$_u$. In GdMn$_6$Sn$_6$, on an E-$k_x$ cut through the D$_1$ kagome Weyl line, we observe the expected linear dispersion, with one cone branch suppressed due to sublattice interference [Fig. 4(a)]. Interestingly, in contrast to the gapless dispersion in GdMn$_6$Sn$_6$, the same cut in Tb$_{0.2}$Gd$_{0.8}$Mn$_6$Sn$_6$ exhibits strongly suppressed spectral weight near the crossing point, indicating the existence of a gap [Figs. 4(b)]. We can directly identify the gap on an energy distribution curve (EDC) through H$_u$ for Tb$_{0.2}$Gd$_{0.8}$Mn$_6$Sn$_6$ [Figs. 4(c), S13 [39]]. Fitting by two Lorentzian peaks yields a gap size $\Delta = (25 \pm 6)$ meV, of the same order of magnitude as the kagome Dirac cone gap observed in $Fe_3Sn_2$ [8]. By contrast, the corresponding EDC for undoped $GdMn_6Sn_6$ is well-fit by a single Lorentzian peak. Corresponding *ab initio* calculations with in-plane magnetic moment exhibit a near-gapless Weyl line [Fig. 4(d)]. While under out-of-plane magnetic order, the Weyl line develops an SOC gap [Fig. 4(e)]. This out-of-plane SOC gap persists along the full Weyl line trajectory, with the largest splitting at $H_u$ [Figs. 4(g)], suggesting enhanced Kane-Mele SOC associated with the out-of-plane magnetic order. Since the kagome Weyl line disperses close to the Fermi level, Berry curvature generated through the Kane-Mele SOC and concentrated along the Weyl line is expected to contribute to an enhanced AHE under out-of-plane magnetism [Fig. S14 [39]]. Taken together, our photoemission spectra and *ab initio* results unambiguously suggest the observation of a magnetically tunable Kane-Mele SOC gap in a kagome Weyl line.

Our systematic ARPES measurements and *ab initio* calculations reveal strongly three-dimensional magnetic Weyl lines originating from the A-A stacked Mn kagome lattice in $(Tb,Gd)Mn_6Sn_6$, which was previously considered to be a quasi-two-dimensional system [4,34]. Moreover, we report, for the first time, that the Weyl line exhibits a strongly magnetization-direction tunable Kane-Mele SOC gap and Berry curvature distribution. Under in-plane magnetic order the gap is negligible, while under out-of-plane order the gap is strongly enhanced, consistent with a leading Kane-Mele SOC contribution. Our results point to the crucial role of magnetization direction and interlayer coupling in the topological electronic structure of $R Mn_6Sn_6$ family and provide a general design principle for nodal lines with tunable mass and Berry curvature geometry. Since the magnetic reorientation transition in $(Tb,Gd)Mn_6Sn_6$ can be tuned to room temperature, the Berry curvature concentrated by the Weyl line may allow the implementation a novel room-temperature anomalous Hall sensor with unusual directional sensitivity [34].

**Acknowledgements:**

The material characterization (ARPES) is supported by the United States Department of Energy (US DOE) under the Basic Energy Sciences program (grant number DOE/BES DE-FG-02-05ER46200). This research used resources of the Advanced Light Source(ALS), a DOE Office of Science User Facility under contract number DE-AC02-05CH11231. Use of the Stanford Synchrotron Radiation Light Source (SSRL), SLAC National Accelerator Laboratory, is supported by the U.S. Department of Energy, Office of Science, Office of Basic Energy Sciences, under contract no. DE-AC02-76SF00515. The authors thank Donghui Lu and Makoto Hashimoto at Beamline 5.2 of the SSRL for support. The authors thank C. Polley, J.Adell and B. Thiagarajan at Beamline Bloch of the Max IV, Lund, Sweden for support. The authors also thank V. Strokov and N. Schröter at Beamline ADRESS of the Swiss Light Source (SLS) at Paul Scherrer Institut, Switzerland for support. The authors also want to thank J. Denlinger at Beamline 4.0.3(MERLIN) of the ALS for support in getting the preliminary data. The authors also acknowledge use of Princeton University's Imaging and Analysis Center, which is partially supported by the Princeton Center for Complex Materials (PCCM), a National Science Foundation (NSF)-MRSEC program (DMR-2011750). I.B. acknowledges the generous support of the Special Postdoctoral Researchers Program, RIKEN during the late stages of this work. T.A.C. acknowledges the support of the National Science Foundation Graduate Research Fellowship Program (DGE-1656466). B. L. acknowledges support from the Alfred P. Sloan Foundation. G.C. would like to acknowledge the support of the National Research Foundation, Singapore under its NRF Fellowship Award (NRF-NRFF13-2021-0010) and the Nanyang Assistant Professorship grant from Nanyang Technological University. T.-R.C. was supported by the Young Scholar Fellowship Program from the Ministry of Science and Technology (MOST) in Taiwan, under a MOST grant for the Columbus Program MOST110-2636-M-006-016, National Cheng Kung University, Taiwan, and National Center for Theoretical Sciences, Taiwan. This work was supported partially by the MOST, Taiwan, grant MOST107-2627-E-006-001. This research was supported in part by Higher Education Sprout Project, Ministry of Education to the Headquarters of University Advancement at National Cheng Kung University (NCKU). M.Z.H. acknowledges support from Lawrence Berkeley National Laboratory and the Miller Institute of Basic Research in Science at the University of California, Berkeley in the form of a Visiting Miller Professorship.

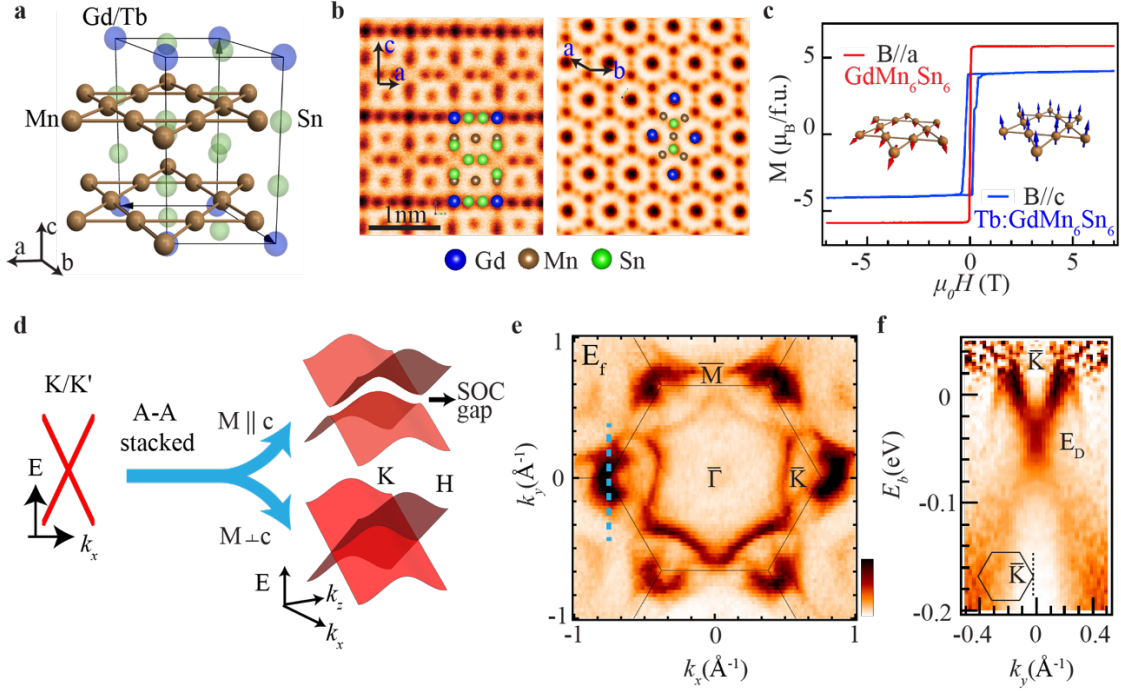

FIG. 1. Structure, magnetism and Dirac crossing in kagome lattice $Tb_xGd_{1-x}Mn_6Sn_6$. (a) Crystal structure, emphasizing the A-A stacked Mn kagome lattice. The black lines: side of the conventional unit cell. (b) Scanning transmission electron microscope (STEM) images of side surface (left panel) and top surface (right panel) of $GdMn_6Sn_6$, where the conventional unit cell is superimposed on top of the STEM results. (c) Magnetic moment verses in-plane applied magnetic field (red line, on $GdMn_6Sn_6$) and out-of-plane magnetic field (blue line, $Tb_{0.2}Gd_{0.8}Mn_6Sn_6$ at T = 10 K. f.u.: formula unit. (d) Illustration of the effect of A-A stacking and magnetization direction: tuning Dirac cone of 2D kagome lattice into nodal line along KH direction with magnetization-direction dependent SOC gap. (e) FS map of $GdMn_6Sn_6$ with (001) termination, taken with 140eV, horizontal polarized light and at 16K. The image is obtained by binning within 10meV around the Fermi level. (f) Renormalized ARPES energy-momentum cut along the dashed line in (e).

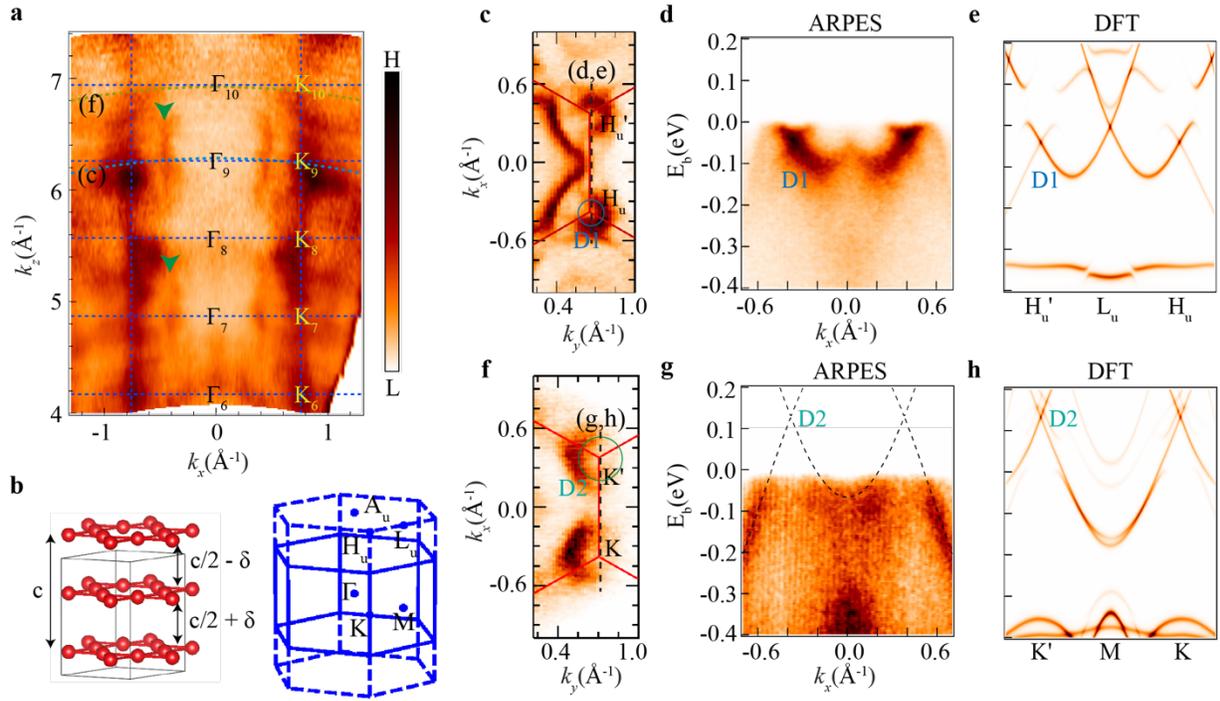

FIG. 2. Photon energy dependence and band unfolding effect of Dirac states. (a) FS map in $k_z - k_x$ plane, showing clear BZ doubling effect of the state marked by the green arrows. Inner potential, $V_0 = 14$ eV. (b) Left: crystal structure with only Mn atoms, which has A-A stacking pattern along c axis with nearly identical interlayer spacing. Right: original BZ (solid line) and doubled BZ (dashed line). (c) and (f) are Fermi surface maps taken at $k_z \approx 9\frac{2\pi}{c}$ and $k_z \approx 10\frac{2\pi}{c}$, respectively. The $k_z$ momentum trajectories of two maps are shown as dashed lines in (a). (d) and (g) are energy-momentum cuts taken along the dashed lines in the corresponding FS maps (c) and (f). The dashed line in (g) is the fitting results using single-orbit kagome model. The two Dirac points are labeled with D1 and D2. (e) and (h) are corresponding theoretical calculations, which shows excellent consistency with (d) and (g), respectively.

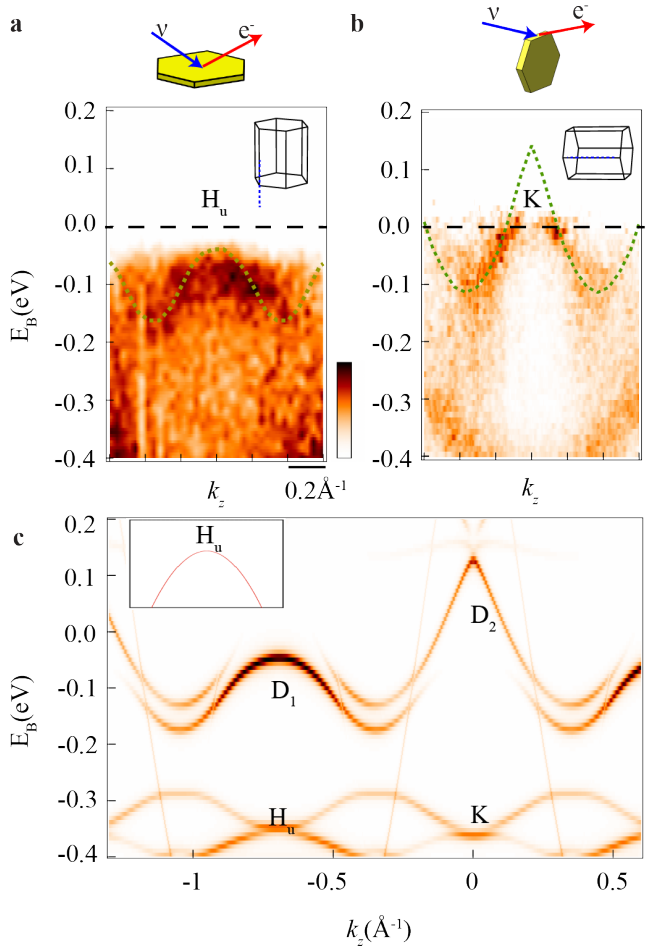

FIG. 3. Strong $k_z$ dispersion of magnetic nodal lines. (a) and (b) are $k_z$ dispersion of the D1 Dirac cones and D2 Dirac cones along K-H direction, respectively. The schematics above the ARPES results show the cleaving plane of each cut. The green dashed lines are the calculation results extracted from (c). The inset in (c) is the zoom-in theoretical band dispersion (without SOC) of D1 Dirac cones near band top, showing the double degeneracy of the nodal line.

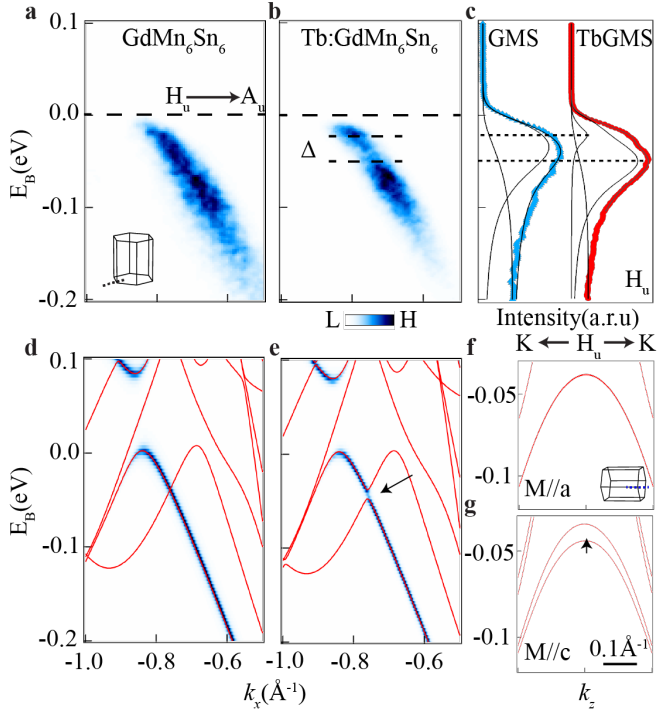

FIG. 4. Magnetic-moment-direction tunable SOC gap of the D1 nodal line. (a) Energy momentum cut along $H_u - A_u$ in the unfolded BZ of GdMn$_6$Sn$_6$. (b) Analogous to (a), but of Tb$_{0.2}$Gd$_{0.8}$Mn$_6$Sn$_6$. (c) EDCs through the crossing point at $H_u$ (also see Fig. S10). (d-f) Calculated Band dispersion with M//a (d) and M//c (e), for comparison with (a-b). The blue lines are unfolded band calculation. (f-g) Calculated Band dispersion along $H_u$-K with M//a (f) and M//c (g). The insets in (a) and (g) denote the direction of the cuts in the unfolded BZ.